\documentclass[11pt]{article}
\usepackage{a4wide}

\usepackage{amsmath, amssymb, amsthm, amsfonts, dsfont, bbm,calc,commath,url,mathtools}
\usepackage{enumerate}
\usepackage{bussproofs}
\usepackage{nicefrac}

\newtheorem{theorem}{Theorem}[section]

\newtheorem{lemma}[theorem]{Lemma}

\newtheorem{corollary}[theorem]{Corollary}

\newtheorem{fact}[theorem]{Fact}

\def\phi{\varphi}

\newcommand{\mathrmL}{{\mathchoice{\mbox{\rm\L}}{\mbox{\rm\L}}{\mbox{\rm\scriptsize\L}}{\mbox{\rm\tiny\L}}}}

\newcommand{\lang}[1]{\ensuremath{\mathcal L}({#1})}

\newcommand{\logic}[1]{\ensuremath{\mathrm {#1}}}

\newcommand\FL[1]{\ensuremath{ \logic{{FL}}_\mathrm{#1} }}
\newcommand\FLew{\FL{ew}}

\newcommand\RPL{\logic{RPL}}

\newcommand\GRPL{\logic{GRPL}}
\newcommand\Luk{\logic{\mathrmL}}

\newcommand\gform[2]{\ensuremath{\langle {#1},{#2} \rangle}}

\newcommand{\alg}[1]{ {\ensuremath{\mathcal {#1}}}}
\newcommand{\Alg}[1]{\ensuremath{\mathbbmss {#1}}}

\newcommand\standardL{\ensuremath{{[0,1]_\mathrmL}}}
\newcommand\rationalL{\ensuremath{{\mathrm{Q}_\mathrmL}}}

\newcommand\standardLQ{\ensuremath{[0,1]_{\mathrmL}^Q}}

\begin{document}
\title{Rational Pavelka logic: the best among three worlds?}
\author{Zuzana Hanikov\'a\\
Institute of Computer Science of the Czech Academy of Sciences\\
182 07 Prague, Czechia\\
email: {\tt hanikova@cs.cas.cz}}
\maketitle

\begin{abstract}
This comparative survey explores  
three formal approaches to reasoning with partly true statements
and degrees of truth, within the family of {\L}ukasiewicz logic. 
These approaches are represented by
infinite-valued {\L}ukasiewicz logic ({\L}), 
Rational Pavelka logic (RPL) and a logic with graded formulas that we refer to as Graded Rational
Pavelka logic (GRPL). Truth constants for all rationals between $0$ and $1$
are used as a technical means to calibrate degrees of truth. 
 {\L}ukasiewicz logic ostensibly features no truth constants except $0$ and $1$; 
Rational Pavelka logic includes constants in the basic language, with
suitable axioms; Graded Rational Pavelka logic works with graded formulas and proofs, 
following the original intent of Pavelka, inspired by Goguen's work. 
Historically, Pavelka's papers precede the definition of GRPL, which in turn precedes RPL;
retrieving these steps, we discuss how these formal systems naturally evolve
from each other, and we also recall how this process has been a somewhat contentious issue in the realm of  {\L}ukasiewicz logic.
This work can also be read as a case study in logics, their fragments, 
and the relationship of the fragments to a logic. 
\end{abstract}

\section{Introduction}
\label{section:intro}

Fuzzy logics with (rational) constants present a \emph{sui generis} research area which 
arises from the preference for greater expressivity of the propositional language. 
The inclusion of constants in the language was an ingenious move that, in retrospect, seems quite natural: 
once it is  admitted that propositions can take many different \emph{truth values}, 
and the idea is embraced that the truth values \emph{form an algebra}, 
the next step---identifying the term-definable functions in the algebra, and introducing
more connectives for some of the functions that are not term-definable in it---proposes itself.
Propositional constants, being nullary connectives, thus present one conceptually easy move 
in this enterprise.\footnote{It is only with the benefit of hindsight that any such conceptual picture could be outlined.
To forestall an (erroneous) impression that the historical development would have followed the arity of the expanding connectives, let us 
recall the early and sophisticated system  of Takeuti and Titani \cite{Takeuti-Titani:Fuzzy},
published as early as 1992, 
employing several families of propositional connectives, including
all rational constants, the  {\L}ukasiewicz connectives, the product conjunction, etc.}

But to be fair, let us take an upfront look at some of the pitfalls of such a move.
The previous paragraph mentions without further specification that the truth values form ``an algebra''.
A typical member of the family of fuzzy logics (say, an extension of the logic $\logic{MTL}$) 
will be algebraizable, and the quasivariety that forms its equivalent algebraic semantics 
will have complicated structure, with many subclasses of independent interest.
This is, indeed, the case of {\L}ukasiewicz logic, whose equivalent algebraic semantics is given by
the variety $\Alg{MV}$ of MV-algebras, which is Q-universal: that is, for every quasivariety 
$\Alg{K}$ of algebras in a finite language, the lattice of
all subquasivarieties of $\Alg{K}$ is a homomorphic image of a sublattice of the lattice of all
quasivarieties of $\Alg{MV}$ \cite{AdamsDziobiak:Q-universal}---hence the subquasivariety lattice is ``as complicated
as it can be'' in a well-defined sense.
By selecting one such algebra---namely, the MV-algebra on the rationals in 
the interval $[0,1]$ with the usual order, introducing a propositional constant for each element of this algebra, 
and introducing axioms that capture the behaviour of the MV-operations on all the elements 
(such axioms are often referred to as the \emph{bookkeeping axioms}
in our \emph{milieu}), one obtains a logic that is still algebraizable but the class forming the equivalent 
algebraic semantics is reduced significantly. In particular,  adding the constants
for rationals in $[0,1]$ with the bookkeeping axioms to $\Luk$ leads to the logic $\RPL$:
the equivalent  algebraic semantics of $\RPL$ is the variety of $\RPL$-algebras,
each nontrivial member of which contains an isomorphic copy of the MV-algebra on the rationals in $[0,1]$, and the class has no nontrivial subquasivarieties \cite[Theorem 9.1]{Gispert-HMS:StrCompWithConstants}.\footnote{But notice that 
the situation is quite different e.g.~for product logic, where the product algebra on the rationals in $[0,1]$
is non-simple, namely has the two-element Boolean algebra as its homomorphic image;
hence any set of axioms for constants that are valid in the product algebra on the rationals is also
valid in the two-element Boolean algebra. 
Thus each product algebra can be expanded to a model of  product logic with rational constants, such that 
the constant indexed with zero is interpreted with $0$ and each constant with nonzero index is interpreted with $1$ (cf.~\cite[Definition 3.7]{Savicky-CEGN:ProductTruthConstants}). 
An analogous situation is not possible in {\L}ukasiewicz logic, where the MV-algebra on the rationals is simple.}

Moreover, the inclusion of constants (or, more generally, any significant expansion of language) turns out to be costly 
when it comes to the possibility of comparing the investigated logic to other systems as to logical strength.
The adherence to logics that rely on the language of classical propositional logic (\logic{CPC}), 
 only enable the formal derivation of \emph{fully true} statements, and are algebraizable with each nontrivial element of the algebraic semantics containing an isomorphic copy of the two-element Boolean algebra, 
places each such logic in question automatically in a broad family of logics
that are comparable in strength to classical logic, or to other significant systems (be it, for example
the family of substructural logics, where indeed many fuzzy logics can smoothly be conceived). 
If the language of a logic is conspicuously
broader that that of $\logic{CPC}$ or other well-known systems,
then one compares \emph{fragments} of that logic to $\logic{CPC}$ or other systems.
As an example of this phenomenon, linear logic with its rich language is often compared to other logics via its fragments (see, e.g., \cite{Galatos-JKO:ResiduatedLattices}
for details). 
Logics with rational constants are ranked naturally by their fragment without constants.

These reservations notwithstanding, fuzzy logics with constants do hold a clear philosophical appeal.
To appreciate it, it is profitable to begin with J.~A.~Goguen's two papers---now 
classics in the area---namely \emph{L-Fuzzy Sets} \cite{Goguen:LFuzzySets} and 
\emph{The logic of inexact concepts} \cite{Goguen:LogicInexactConcepts}.\footnote{Notice, for example, that \cite[Theorem 2.1.8]{Hajek:1998} refers to the residuum in the standard product algebra as \emph{Goguen implication}.}
Both these works tailed closely Zadeh's work \emph{On fuzzy sets} \cite{Zadeh:1965}, 
pinpointing the kind of uncertainty they aim to address (Goguen in \emph{L-Fuzzy Sets} speaks of
\emph{ambiguity} and seeks to differentiate it from probability),
and developed
the idea of algebra of truth values, starting from the notion of a complete lattice 
and adding a multiplication $\odot$ that left- and right-distributes over arbitrary joins (possibly with additional properties,
such as commutativity for $\odot$ or distributivity of the lattice operations). 
Goguen's considerations thus lead to the notion of a \emph{residuated lattice}, first considered by Ward and Dilworth
\cite{Ward-Dilworth:ResiduatedLattices}.
Thus he significantly contributed to the development of \emph{algebraic fuzzy logic}, which was later on pursued by other researchers\footnote{In particular, see \cite[Section 2.2]{Belohlavek:PavelkaRetrospectProspect} for a discussion of the outstanding merit of Goguen's analysis.
}: see the three volumes of the \emph{Handbook of Mathematical Fuzzy Logic} \cite{HandbookMFL-IandII,HandbookMFL-III} and the references therein for general overview, 
and in particular \cite{Pavelka:Fuzzy,Hajek:FuzzyLogicPointOfView,
Novak:CompletenessFirstOrder,Hajek:ArHierarchyI,Hajek:1998,Hajek-Paris-Shepherdson:Pavelka,Cignoli-Esteva-Godo:LukasiewiczConstants,Novak:CountableSyntaxRevisited,
Belohlavek:PavelkaRetrospectProspect,Hanikova:ImplicitDefConstLuk,Hanikova:ComplexityValDegree} 
for expansions of {\L}ukasiewicz logic with constants; more generally 
\cite{Esteva-Godo-Marchioni:FuzzyLogicsEnriched,Cintula:noteAxiomatizationsPavelka} 
for expansions of wider classes of  logics with propositional constants.
Furthemore the survey \cite{Novak:Handbook} and the references therein 
complete the picture for logics with graded formulas.

Taking into account the eventual impact of Goguen's work,
we can say that fuzzy logic was historically born with the capability of formally processing partly true
statements and plausibly estimating the validity of derivations
of such statements. 
Logics with rational constants can naturally be viewed as continuing this research line and 
enhancing this ability, within the framework of truth preserving logics;
see \cite[Section 2]{Esteva-Godo-Marchioni:FuzzyLogicsEnriched} 
and references therein for a discussion of \emph{logics preserving degrees of truth},
another and perhaps more radical departure from the tradition of deriving true conclusions from true axioms.  

Goguen advocated algebraic computations with truth degrees, but he did not use explicit truth constants.
This was first done by Pavelka \cite{Pavelka:Fuzzy}, who continued the general line of building the logic 
(taken as a structural consequence relation) of a complete residuated lattice, and exemplified his general treatment with building 
the calculus of the infinite-valued {\L}ukasiewicz logic with constants for all the reals. 
After authoring the series of three papers (which formed the subject matter of his dissertation),
Pavelka did not continue this research. It was taken up by H\'ajek and by Nov\'ak, in two rather different
attitudes, the evolution of  both of which will be discussed in Section 2.

While this survey, at various points, considers many different systems of fuzzy logic with constants, 
it aims to compare three particular systems, taken as three representatives of logics of comparative truth. 
These three systems, introduced in sufficient detail below, are as follows:
\begin{itemize}
\item infinite-valued {\L}ukasiewicz logic $\Luk$ as in \cite{Lukasiewicz-Tarski:Untersuchungen,
Rose-Roser:Fragments} and in the monographs \cite{Cignoli-Ottaviano-Mundici:AlgebraicFoundations, Mundici:Advanced};  
\item {Rational Pavelka logic} $\RPL$ as in \cite{Hajek:FuzzyLogicPointOfView, Hajek:1998};
\item Graded Rational Pavelka logic $\GRPL$ as in \cite{Hajek:ArHierarchyI}.
\end{itemize}

The evolution of the logics is outlined in sufficient detail in Section \ref{section:evo},
while Section \ref{section:formally} gives a technical background to make this survey reasonably self-contained.
Section \ref{section:relationship} provides ways of obtaining 
information on the provability relation of one of the logics
from the provability relation of another one. 
It was in order to enable such a comparison that we chose two logics ($\RPL$ and $\GRPL$) 
each of which expands the infinite-valued {\L}ukasiewicz logic
with constants for the \emph{rationals} rather than the \emph{reals}, despite the fact that the research school
that investigates the logics with graded formulas puts more emphasis on languages where the constants
represent all the reals. Such a logic in turn would correspond to a \emph{Real Pavelka logic} in place of our $\RPL$,
for the sake of a comparison.
Finally Section \ref{section:discussion} discussed some views on the presented material, including that of the author,
and Section \ref{section:conclusion} concludes the survey.

The technical aspects of the relationship between $\Luk$, $\RPL$, and $\GRPL$ 
are either published or implicit in existing literature.  
Technical presentations of various authors may even casually throw in a remark on the author's preference of one logic
over the others, even though the logics have existed side by side for decades. 
Our aim here is to highlight the very tight relationship of the three systems, but
perhaps especially of $\RPL$ and $\GRPL$,
the broader impact of which  does not appear to have been made explicit.
Since the argument that inspired this work is about propositional logic, we remain 
throughout on the propositional level, where the ideas and solutions can clearly be exposed.

\section{Evolution of the three  logics}
\label{section:evo}

{\L}ukasiewicz infinite-valued logic was first considered by Lukasiewicz and Tarski in their paper 
\cite{Lukasiewicz-Tarski:Untersuchungen}, 
published in 1930. This logic, along with its finite-valued counterparts\footnote{which emerge, nowadays, as axiomatic extensions of the (weaker) infinite-valued logic}, prominently
the three-valued system introduced by {\L}ukasiewicz already in 1920 \cite{Lukasiewicz:ThreeValued}, 
sought to capture some types of uncertainty in the semantics of natural language  that users have little difficulty internalizing and employing
on a daily basis, but that appeared difficult to model within---in fact appeared inconsistent with---classical logic.

The pioneering work of {\L}ukasiewicz gives some leeway in the interpretation of a 
\emph{third truth value}, as introduced in \cite{Lukasiewicz:ThreeValued}: 
it is described as \emph{indifferent} in \cite{Lukasiewicz:onDeterminism}. 
The latter retrospective work aims at recording {\L}ukasiewicz's early account of 
his deep-rooted doubt on the validity of the law of the excluded middle, and can also be read as 
 \emph{prolegomena} to his later systems of modal logic,
as well as (or indeed primarily as) to his formal, truth-functional systems where the law of the excluded middle fails.\footnote{
Interestingly, {\L}ukasiewicz's views on the very core tenets of philosophy of logic
apparently underwent a change, engendered at least in part by his work in many-valued and modal
logic; e.g., towards the end of his career he would no longer admit a distinction
 between  empirical and a priori sentences. See \cite{Rybarikova:LukasiewiczQuineEmpApriori} for details. 
}
These initial considerations developed into one of several grand avenues toward formal many-valued and fuzzy logic (with other such considerations
provided by Zadeh and Goguen, G\"odel, Heyting, or Post; see \cite[Section 10.1]{Hajek:1998}) 
and \emph{a fortiori} also to, on the one hand, advanced theory of {\L}ukasiewicz logic and MV-algebras---see 
\cite{Cignoli-Ottaviano-Mundici:AlgebraicFoundations, Mundici:Advanced, DiNola-Leustean:HandbookMValgebras} for overviews, 
and on the other, applications and (re-) interpretations that the formal system of fuzzy logic has to offer in the area
of reasoning, broadly conceived: here we cannot aim for a complete picture, see e.g. \cite{Kroupa-Teheux:ModalLukCoalitionPower,Bilkova-FritellaMajerNazari:BeliefBasedInconInf,Cintula-GNS:DegreesToEleven} for some  recent developments.

Taking $\neg$, $\to$, and (definable) $\vee$ as basic connectives, {\L}ukasiewicz and Tarski listed a set of axioms, conjectured earlier by {\L}ukasiewicz to be complete w.r.t. the infinite-valued semantics (when modus ponens is taken as the only rule of deduction).
Rose and Rosser \cite{Rose-Roser:Fragments} proved the standard completeness of this axiomatization, 
while Hay \cite{Hay:Axiomatization} extended it both to a propositional finite strong standard completeness proof, and to a first-order axiomatization 
with an infinitary rule 
(the set of first-order tautologies is not recursively enumerable \cite{Scarpellini:PredikLukNonComplete} and is actually $\Pi_2$-complete \cite{Ragaz:PhD}).

A powerful impulse for the development of {\L}ukasiewicz logic came 
with the works of Chang (see \cite{Chang:MVAlgebras,Chang:ANewProof}) on the algebraic semantics
provided by the class of MV-algebras. 
This work provided what nowadays would be referred to as the equivalent algebraic semantics for the logic, 
and MV-algebras have since gathered
a voluminous and manifold body of research work (cf.~the already mentioned \cite{Cignoli-Ottaviano-Mundici:AlgebraicFoundations, Mundici:Advanced, DiNola-Leustean:HandbookMValgebras}; proof theory 
of {\L}ukasiewicz logic has been covered in \cite{Metcalfe-Olivetti-Gabbay:ProofTheoryFuzzy}).

Goguen's main contribution to the evolution of fuzzy logic consists in adopting an \emph{axiomatic
approach} to establising an algebraic structure on the range of Zadeh's fuzzy sets (represented as, 
or downright identified with, their characteristic functions),
and on the collection of fuzzy sets as such, and quickly proceeding to 
identifying an appropriate class of algebras for this enterprise. 
In \cite[p.~147]{Goguen:LFuzzySets} he writes: 
``We have used the axiomatic method, in the sense that our underlying
assumptions, especially about $\alg{L}$, are abstract; it can thus be ascertained to
what extent our results apply to some new problem.'' 
The subsequent \cite[p.~355]{Goguen:LogicInexactConcepts} continues on this note:
``Our general method will be to consider a fixed but arbitrary truth set $\alg{L}$, and to deduce
properties of the logic using only the closg axioms.
This has two advantages: \emph{generality}; and \emph{invariance}.
Anything which follows from the closg axioms is true for any particular closg; and also results will be invariant under
closg automorphisms.''
In particular, in these two papers the acronym \emph{closg} stands for a \emph{complete lattice-ordered semigroup}:
since the semigroup operation is required to distribute over infinite joins, 
this gives left and right residuation. In \cite{Goguen:LogicInexactConcepts}, 
Goguen adopts an axiomatics that in modern parlance would amount to a completely ordered
$\mathrm{FL}_{\mathrm{w}}$-algebra.\footnote{The commutativity of the semigroup operation
is not strictly imposed, but it is implicitly used throughout the paper.}
Both Goguen's papers maintain a non-technical facet with a plethora of practical examples that make
clear the width of the author's scope, ranging from engineering considerations to semantical analysis;
and yet both papers read as cornerstones in the philosophy of vagueness. 

Furthemore, Goguen \cite{Goguen:LogicInexactConcepts}  advocates computations on truth degrees of formulas as a counterpart of formula derivation, 
allowing to estimate algebraically the validity of conclusions in chains of formal derivations.
This is Goguen's blueprint for truth constants, elaborated and deployed subsequently by Pavelka.
On p.~365 he remarks on his approach as follows:
``A traditional way to develop a logical system is through its tautologies.
[\dots] but no list of tautologies can encompass the entire system because we want to perform
calculations with degrees of validity between $0$ and $1$. In this sense, the logic
of inexact concepts does not have a \emph{purely} syntactic form. Semantics, 
in the form of specific truth values of certain assertions, is sometimes required.'' 
He suggests that proofs in his logic capturing vagueness
may be of the form $[A\Rightarrow B]\geq a$, with $A,B$ well-formed formulas of 
a fixed language  and $a$ a value from a fixed algebra, representing a lower bound on the validity
of the derivation $A\Rightarrow B$; the latter can be computed from a given derivation.
The operator $[]$  assigns to each proposition $P$ its
truth value $[P]$ in a closg (cf.~page 333 of the paper). However, the expression $[P]$ is 
not, in Goguen's work, an element of the syntax.
These two papers of Goguen are both broader in scope and richer in detail than the brief account here suggests, but
hopefully it does do enough to recall how his work posed the challenge to develop a formal calculus for propositions that naturally admit
truth values other than the boolean ones, and at the same time indicated a way towards such a calculus.

Pavelka starts by extending the notion of a consequence relation to  \emph{fuzzy theories} (that is, 
fuzzy sets of formulas; again the range is assumed to be a complete lattice $\alg{L}$, possibly expanded with other operations)
and then embarks on an investigation of conditions under which one can obtain a corresponding 
syntactical consequence relation. This starts from the definition of an $\alg{L}$-syntax, consisting
of an $\alg{L}$-fuzzy set of axioms and $\alg{L}$-rules of inference; this explicitly involves 
names for elements of $\alg{L}$,
although in the general form (as presented in \cite{Pavelka:Fuzzy-I}, for example) it is by no means clear 
that these names, which take explicit part in syntactic derivations, need
also to feature in the set of formulas $F$ as nullary connectives.  

Part II of his work \cite{Pavelka:Fuzzy-II} is a detailed study of the properties of algebraic structures that provide the \emph{truth values}
of his logic. Here, Pavelka uses the notion \emph{residuated lattice} in the sense of a $\mathrm{FL}_{\mathrm{ew}}$-algebra, 
following Ward and Dilworth \cite{Ward-Dilworth:ResiduatedLattices} and Goguen \cite{Goguen:LogicInexactConcepts}.
In the last paper of the series \cite{Pavelka:Fuzzy-III}, Pavelka deduces that the only 
eligible candidate for a complete axiomatization of the fuzzy consequence relation, considering the residuated lattices on $[0,1]$,
is the pair of {\L}ukasiewicz logic and the algebra $\standardL$, and proves the famous Pavelka completeness.
Still, a major part of the paper \cite{Pavelka:Fuzzy-III} 
the construction is kept quite general (e.g. considering also additional operations) 
and only for the final part 
the exposition mentions concrete algebras, such as the standard MV-algebra or its finite subalgebras.
See also B\v{e}lohl\'avek's account in \cite[Section 3]{Belohlavek:PavelkaRetrospectProspect}.

Pavelka's work inspired the research area of logics with graded syntax,
whose detailed inspection is beyond the scope of this paper and we limit the exposition
to a few brief remarks; see \cite{Novak:Handbook} for a full account.
Similarly as in the case of the logic $\Luk$ above, our intent here is not to capture
every work ever published on the system, but to outline relevant aspects of its evolution.
Nov\'ak established completeness for first-order version
of Pavelka's system in \cite{Novak:CompletenessFirstOrder} (this seems to have been anticipated
at the end of \cite{Pavelka:Fuzzy-III}); the logic introduces constants for the reals 
into the system of infinite-valued {\L}ukasiewicz logic.  
One milestone of the development of graded syntax logics was the publication \cite{Novak-Perfilieva-Mockor:2000}.
Thereafter, some advanced topics, such as a fuzzy type theory \cite{Novak:FTT} were pursued.

From a model-theoretic point of view, there is nothing extraordinary 
in adding constant elements, the \emph{names} for each element of a structure, 
to a formal language; the move is familiar from the construction of the diagram of a structure.
It is equally clear that going beyond countably many constants, 
while necessitated by the choice of the structure, can be viewed as an aberration 
from what is commonly understood by the term
\emph{reasoning}, since it makes the latter non-finitary and thus in particular,
beyond the reach of commonplace reasoners---humans, or even machines.

The gap between logics with constants for the reals (or for elements of other complete
residuated lattices) and those logics that introduce constants for the rationals or other countable
subdomains of $[0,1]$ was first bridged in  H\'ajek's early paper \cite{Hajek:ArHierarchyI},
which keeps the reals as truth values of propositions, but reduces the set of propositional constants,
the set of grades occurring in graded formulas, and grades of membership of formulas in fuzzy theories
to the rationals, while still proving Pavelka completeness (see below).
This paper  keeps the graded elements in the syntax; but  the year 1995 
when it was published saw H\'ajek introduce, in a plenary lecture at the SOFSEM conference and the attending
paper  \cite{Hajek:FuzzyLogicPointOfView} in the proceedings, the system $\RPL$ for the first time. 
It was later presented in more detail and with proofs 
in \cite{Hajek:1998}.\footnote{On countable sets of grades in the graded syntax milieu, 
see  \cite{Novak:CountableSyntaxRevisited}.}

\section{Introducing the logics formally}
\label{section:formally}

This section provides a brief exposition of the  Hilbert deduction 
systems available for the three
logics $\Luk$, RPL, and GRPL, along with their algebraic semantics.  
The material is selected with the  prospect of a comparative study.
Comprehensive surveys of {\L}ukasiewicz logic include the volumes \cite{Cignoli-Ottaviano-Mundici:AlgebraicFoundations, Mundici:Advanced} and the chapters \cite{DiNola-Leustean:HandbookMValgebras} 
and \cite[Chapter 3]{Hajek:1998}; see also the references therein.
{\L}ukasiewicz logic is semantically motivated; 
as already remarked, the expansion with constants for rationals adds to its dependence on the semantics.
Accordingly, we provide the rudiments of
the theory of MV-algebras, the equivalent algebraic semantics of {\L}.

{\L}ukasiewicz logic is arguably quite strong. One hallmark of this is the fact
that its connectives are interdefinable, whereby it is possible to opt for only a few 
basic ones and define the remaining ones, out of the usual set of connectives
considered for this logic,
which comprises constants $0$ and $1$, unary $\neg$ (negation), 
and binary $\odot$ (strong conjunction or multiplication), $\oplus$ (strong disjunction),
$\land$ and $\lor$ (lattice conjunction and disjunction), $\to$ (implication), and $\equiv$ (equivalence).
One tradition, following the Hilbert formal system as introduced by {\L}ukasiewicz and Tarski, 
takes the set $\{\neg, \to\}$ for basic connectives. We shall refer to this language as $\lang{\Luk}$.
Then one defines other connectives as shortcuts:
$\alpha\vee\beta$ as $(\alpha\to\beta)\to\beta$; $\alpha\oplus\beta$ as $\neg\alpha \to\beta$;
$\alpha\odot\beta$ as $\neg(\neg\alpha\oplus\neg\beta)$; $\alpha\wedge\beta$ as $\alpha\odot(\alpha\to\beta)$;
and $\alpha\equiv\beta$ as $(\alpha\to\beta)\wedge(\beta\to\alpha)$. 
By recursive application of these metarules, each formula in the full language $\lang{\Luk}$ can
be rewritten as a formula in the basic language.
Further shortcuts may include $\varphi^n$, standing for $\overbrace{\phi\odot\ldots\odot\phi}^{n\mbox{ times}}$,
and $n\cdot \varphi$, standing for $\overbrace{\phi\oplus\ldots\oplus\phi}^{n\mbox{ times}}$.

The algebraic tradition usually opts for $\{\neg, \oplus\}$ as the set of basic function symbols\footnote{Then
$\to$ is introduced as $\neg\alpha\oplus\beta$ and the remaining definitions are as before.},
on the ground of a tight connection between MV-algebras and lattice-ordered abelian groups (see below).
Chang's MV-algebras \cite{Chang:MVAlgebras,Chang:ANewProof} were introduced as a tool for algebraic study of {\L}ukasiewicz logic and 
can in fact be shown to be the equivalent algebraic semantics of $\L$ in the sense of Blok and Pigozzi 
\cite{Blok-Pigozzi:AlgebraizableLogics}. 
Our definition presents MV-algebras as a subclass of $\FLew$-algebras. 
An MV-algebra is an algebraic structure of the form 
$\alg{A} = \langle A,\odot,\to, \wedge, \vee, 0,1 \rangle$  
where $\langle A, \wedge, \vee, 0, 1 \rangle$ 
is a bounded lattice, $\langle A, \odot, 1 \rangle $ is a commutative monoid with unit $1$, 
$\odot$ and $\to$ form a residuated pair ($x\odot y\leq z$ if and only if $x\leq y\to z$ for $x,y,z\in A$), 
and moreover for $x,y\in A$ we have $(x\to y)\lor (y\to x)=1$ (semilinearity),  
 $x\wedge y = x\odot (x\to y)$ (divisibility), and $ (x\to 0)\to 0=x$ (involutiveness).
MV-algebras can be shown to form a variety.
The top and bottom element of any nontrivial MV-algebra $\alg{A}$ form a two-element 
Boolean subalgebra of $\alg{A}$; in particular, the variety of Boolean algebras is a subvariety
of $\Alg{MV}$. A totally ordered MV-algebra is called a \emph{chain};
it can be shown that every MV-algebra is a subdirect product of chains, and hence,
the lattice reduct is distributive.
Another fact that contributes to the informal claim
on strength of $\logic{\Luk}$ is the extant complete description 
of the lattice of its axiomatic extensions \cite{Komori:SuperLukasiewiczPropositional}, which is 
dually isomorphic to the lattice of subvarieties of $\Alg{MV}$.

Examples of MV-algebras can be obtained from $\ell$-groups (lattice-ordered abelian groups).
Let $\alg{G}=\{G,\wedge,\vee,+,-,0\}$ be an $\ell$-group.
Let $u$ be a positive element of $G$.
For each $x,y\in[0,u]$, define
$x\oplus y = u\land (x+y)$
and $\neg x = u-x$ (the operations on the right side of equations are the $\ell$-group operations).
Then $\langle [0,u],\oplus,\neg,0\rangle$ is an MV-algebra, 
denoted $\Gamma (\alg{G},u)$.
In fact, there is a categorical equivalence between the category of MV-algebras and the category 
of $\ell$-groups with a strong unit, obtained by Mundici \cite{Mundici:AbelianLGr}.
Prior to that, Chang \cite{Chang:ANewProof} established a correspondence between totally ordered $\ell$-groups 
and MV-chains. 

The intended semantics for $\L$, often referred to as the \emph{standard} MV-algebra
and denoted $\standardL$, can be simply introduced as 
 $\Gamma (\mathds{R}, 1)$ (where $\mathds{R}$ stands for the additive $\ell$-group
on the real numbers).
Its subalgebra on the rational numbers will be denoted $\rationalL$.
Moreover, the finite subalgebra of $\standardL$ (and of $\rationalL$) on the domain
$\{0,\nicefrac{1}{n},\dots,\nicefrac{n}{n}\}$ is denoted $\mathrmL_{n}$;
for $n=1$, one just obtains the two-element Boolean algebra.

 A function $f\colon [0,1]^n \rightarrow [0,1]$ is a \emph{McNaughton function} provided that it is continuous, 
 piecewise linear (there are finitely many linear polynomials $\{p_i\}_{i\in I}$,
with $p_i(\bar x) = \Sigma_{j=1}^n a_{ij}\,x_j + b_i$, 
such that for any $\bar x\in [0,1]^n$ there is an $i\in I$ with $f(\bar x)= p_i(\bar x)$), and    
 the polynomials $p_i$ have integer coefficients $\bar {a_i}, b_i$.
The paper \cite{McNaughton:FunctionalRep} established what is known as McNaughton theorem:
namely, McNaughton functions {\it concide} with 
 term-definable functions in $\standardL$.
It is clear from this characterization that no constant function is term-definable
except for the constant $0$ and the constant $1$.

The following are axiom schemata of {\L}ukasiewicz logic presented in \cite{Lukasiewicz-Tarski:Untersuchungen}\footnote{The original
set included also the axiom $(\varphi\to\psi)\vee(\psi\to\varphi)$, derivable from the rest.}:
\begin{itemize}
\item[(A1)] $\varphi\to(\psi\to\varphi)$;
\item[(A2)] $(\varphi\to\psi)\to((\psi\to\chi)\to(\varphi\to\chi))$;
\item[(A3)] $((\varphi\to\psi)\to\psi) \to ((\psi\to\varphi)\to\varphi)$;
\item[(A4)] $(\neg\varphi\to\neg\psi) \to (\psi\to\varphi)$.
\end{itemize}
The  deduction rule is modus ponens: $\varphi, \varphi\to\psi \triangleright \psi$.
{\L}ukasiewicz logic, then, is formally identified with the provability relation $\vdash_{\mathrmL}$
for these axioms and rule (it follows that the logic is finitary).

The \emph{finite strong standard completeness theorem} for $\L$ states that
if $\Gamma$ is a finite set of formulas and $\phi$ a formula of the language $\lang{\Luk}$,
then $\Gamma\vdash_{\Luk}\phi$ if and only if $\Gamma\models_{\standardL}\phi$ (this result is implicit in 
\cite{Chang:ANewProof}; see also \cite[Lm.\ B]{Hay:Axiomatization}, the discussion in 
\cite{Gispert-Torrens:QuasivarietiesSimpleMV}, and \cite [Lm.~3.2.11]{Hajek:1998}). 
The result is equivalent to stating that the variety of MV-algebras 
is generated by its member $\standardL$ as a 
quasivariety.

\medskip

\emph{Rational Pavelka logic} ($\RPL$) was first presented in 
\cite{Hajek:FuzzyLogicPointOfView}, and subsequently in much more detail in \cite[Section 3.3]{Hajek:1998},
as a simplified variant of Pavelka's system \cite{Pavelka:Fuzzy}.
$\RPL$ expands the language $\lang{\Luk}$ with a set
 ${\cal{Q}} = \{ \bar{q} \mid q\in \mathds{Q}\cap[0,1]\}$ of constants;
this expanded language will be referred to as the language of $\RPL$ ($\lang{\RPL}$). 
The \emph{bookkeeping axioms} are the formulas
\begin{itemize}
\item  $\overline{q} \to \overline{r} \equiv \overline{q \to^{\mathrmL} r}$;
\item $\neg \overline{q} \equiv \overline{ \neg^{\mathrmL}\, q }$;
\item $\overline 1 \equiv 1 $;
\item $\overline 0 \equiv 0$ 
\end{itemize}
for all rationals $q$ and $r$; on the right-hand side, $\to^\mathrmL$ denotes the residuum operation of 
$\standardL$ and $\neg^\mathrmL$ denotes the negation thereof, for the sake of clarity;
thus, e.g., the display  $ q \to^{\mathrmL} r$ denotes the rational obtained 
by applying the binary operation $\to^{\mathrmL}$ to the rational numbers $q$ and $r$ in $[0,1]$, and 
$ \overline{ q \to^{\mathrmL} r } $ denotes the constant indexed by that rational. 

Notice that the bookkeeping axioms are obtained in a uniform way, given the set of basic connectives 
of the presentation of the logic $\Luk$ (here in particular, for the language $\lang{\Luk}=\{\to, \neg\}$),
provided that such a set of connectives guarantees that each McNaughton function is term definable.
In general, the set of bookkeeping axioms consists of all formulas
$\circ(\overline{q_1},\dots,\overline{q_n}) \equiv \overline{q_{\circ^\mathrmL(q_1,\dots,q_n)}}$ 
for each basic connective $\circ$.
Given these axioms, analogous bookkeeping formulas can then be proved also for any of the defined connectives.

Again $\RPL$ is an algebraizable logic (see, e.g., the discussion in \cite{Esteva-GGN:AddingTruthConstants})
with the the class of $\RPL$-algebras forming its equivalent algebraic semantics:
an $\RPL$-algebra $\alg{A}$ is an algebra in the language  $\lang{\RPL}$ (i.e., interpreting the language of MV-algebras and
all the rational constants from ${\cal Q}$), such that the $\lang{\Luk}$ reduct of $\alg{A}$ 
is an MV-algebra, and all of the bookkeeping axioms are valid in $\alg{A}$.
Moreover, $\standardLQ$ denotes the expansion of $\standardL$ with constants from   ${\cal{Q}}$
where each rational constant $\overline{r}$ 
is interpreted with $r$.

The usual strong completeness theorem w.r.t.~chains holds: for a set $\Gamma\cup\{\phi\}$ of formulas of $\lang{\RPL}$
we have $\Gamma\vdash_{\RPL}\phi$ if and only if, for each $\RPL$-chain $\alg{A}$, 
$\Gamma\models_{\alg{A}}\phi$  (see, e.g., \cite{Esteva-GGN:AddingTruthConstants}).
Also the following theorem has been obtained: 

\begin{fact} {\bf (Finite strong standard completeness theorem for $\RPL$)} {\rm \cite[Theorem 3.3.14]{Hajek:1998}} 
If $\Gamma\cup\{\phi\}$ is a finite set of formulas in the language $\lang{\RPL}$, then 
$\Gamma\vdash_{\RPL}\phi$ if and only if $\Gamma\models_{\standardLQ}\phi$.
\end{fact}

In the algebraization of $\RPL$, the translation of propositional formulas to algebraic identities
(denoted $\tau$ in \cite{Galatos-JKO:ResiduatedLattices}), given the axiom $\overline{q} \to \overline{r} \equiv \overline{q \to^{\mathrmL} r}$,
yields $\overline{q} \to \overline{r} \equiv \overline{q \to^{\mathrmL} r} \approx 1$, 
which is equivalent in each $\RPL$-algebra
to $\overline{r} \to \overline{s} \approx \overline{r\to^{\mathrmL}s}$.
Analogously we obtain $\neg \overline{q} \approx \overline{ \neg^{\mathrmL}\, q } $ and 
$\overline 1 \approx 1$ and $\overline 0 \approx 0$ from the other types of axioms of $\RPL$.
This set of identities entails the \emph{equational diagram} of the algebra $\rationalL$ (see e.g. \cite{sep-modeltheory-fo}).
Indeed, our expansion of the language 
$\lang{L}_{\mathrmL}$ to the language of $\RPL$ is exactly with constants for the elements of 
$\rationalL$, and it is not difficult to see that the bookkeeping axioms imply all atomic sentences
and their negations  involving only rational constants 
that are true in $\rationalL$, and hence
in $\standardL$. For identities this is easily observed using induction on term structure, for negated identities it is useful to realize that
any model of the bookkeeping axioms contains a homomorphic image of $\rationalL$, and since
this algebra is simple, it contains an isomorphic copy of $\rationalL$. 
As remarked in Section \ref{section:intro},
this can be seen as a considerable drawback of $\RPL$:
the expansion with  constants, aimed at increasing expressivity, 
restricts severely the class that forms its equivalent algebraic semantics, compared to the logic $\Luk$.

The apex of Pavelka's treatment of degrees of truth is a completeness theorem for partial truth.
For any set of formulas $\Gamma\cup\{\phi\}$ in the language $\lang{\RPL}$, 
the \emph{provability degree} of $\phi$ over $\Gamma$
is defined as 
$$ |\phi|_\Gamma = \sup \{ r \mid \Gamma \vdash_{\RPL} \overline{r} \to \varphi  \}$$
whereas, in the same setting, the \emph{validity degree} is 
$$ \|\phi\|_\Gamma  = \inf \{ v(\phi) \mid v(\psi)=1 \mbox{ for each } \psi\in\Gamma\}$$
where $v$ denotes assignments in $\standardLQ$.

\begin{fact} {\bf (Pavelka completeness theorem for $\RPL$)} {\rm \cite[Theorem 3.3.5]{Hajek:1998}}
For $\Gamma\cup\{\phi\}$ a  set of formulas in the language $\lang{\RPL}$,
$$|\phi|_\Gamma = \|\phi\|_\Gamma.$$
\end{fact}

\medskip


A system that will be  called \emph{Graded Rational Pavelka Logic} ($\GRPL$)
for the scope of this paper was introduced in \cite{Hajek:ArHierarchyI};
we follow that presentation (see also the references therein). 
This system can be viewed as an early step, taken by H\'ajek, in further developing Pavelka's work,
taking into account also the work of Nov\'ak.  
Notice in particular that, contrary to our presentation here, $\GRPL$ precedes $\RPL$ chronologically. 
$\GRPL$ simplifies considerably the works of Pavelka in that only rational
truth constants are used while still obtaining Pavelka completeness.
Moreover, H\'ajek also  simplifies the axioms, pointing out, among other things, that
Pavelka was presumably not familar with the work of Rose and Rosser \cite{Rose-Roser:Fragments}.
The system $\GRPL$ admits graded formulas and fuzzy theories with graded proofs, 
iconic of Pavelka's work and (at least ostensibly) absent from $\RPL$ as presented
in \cite{Hajek:1998}.

The language of $\GRPL$ coincides with that of $\RPL$.\footnote{As a matter of fact
H\'ajek  in \cite{Hajek:ArHierarchyI} starts out with constants for all reals but
very soon switches to employing rational constants only. This appears to be a precondition 
for some of the  results in his paper \cite{Hajek:ArHierarchyI}, which are computational:
 determining the position of $\GRPL$ in
the arithmetical hierarchy is the main technical achievement.
It is therefore essential that the strings representing the formulas, constants included,
be finite, and hence amenable to algorithmic considerations.}
A \emph{fuzzy theory} $\Gamma$ is a fuzzy set of  formulas\footnote{This definition, via Pavelka,
draws on Zadeh's paper on fuzzy sets; also the ontology offered there for fuzzy sets is maintained---e.g., 
a fuzzy set $S$ is a \emph{function} (here, on the domain of formulas), so the
membership of an object $m$ in the set $S$ is denoted $S(m)$.}
 in the language of $\RPL$:  
that is, each formula $\psi \in \lang{\RPL}$ belongs 
to $\Gamma$ in some grade, which is a \emph{rational} number, denoted $\Gamma(\psi)$.
An assignment $v$ respects $\Gamma$ provided that $\Gamma(\alpha) \leq v(\alpha)$ 
for each $\alpha\in \lang{\RPL}$. 
The \emph{validity degree} of $\phi$ in $\Gamma$ is 
$\inf \{ v(\phi) \mid v\mbox{ respects }\Gamma\}$; if $r$ is the validity degree of
$\phi$ in $\Gamma$, one can write $\Gamma \models_r \phi$.  

A graded formula is a pair $\gform{r}{\varphi}$,  
with $r$ a rational (called the \emph{grade} of $\varphi$) and $\varphi$ any formula in $\lang{\RPL}$. 
The following list specifies the fuzzy set of axioms  of $\GRPL$:
\begin{itemize}
\item[(A)] Instances of the axioms of $\Luk$ in grade $1$;
\item[(B)] the constant $\overline{q}$ in grade $q$, for each rational $q\in [0,1]$;
\item [(C)] $\neg \overline{q} \equiv \overline{\neg^{\mathrmL} q}$ and $\overline{q} \to \overline{r} 
	\equiv \overline{q\to^{\mathrmL} r} $ in grade 1, for all rational $q,r\in [0,1]$ {\bf (bookkeeping)}
\item [(D)] all other formulas of $\lang{\RPL}$ in grade $0$.
\end{itemize} 
The rules of deduction  are 
\begin{itemize}
\item [(E)] 
\AxiomC{$\gform{q}{\varphi}$}
\AxiomC{$\gform{r}{\varphi\to\psi} $}
\BinaryInfC{$\gform{q\cdot^\mathrmL r}{\psi}$}
\DisplayProof \ \
 for all rational $q,r \in [0,1]$ {\bf (graded modus ponens)}\footnote{see \cite{Goguen:LogicInexactConcepts}};
\item [(F)] 
\AxiomC{$\gform{q}{\varphi}$} 
\UnaryInfC{$ \gform{r\to^\mathrmL q}{\overline{r}\to\varphi}$} 
\DisplayProof
\ \ for all rational $q,r\in[0,1]$ {\bf (lifting)}. 

\end{itemize}
A graded proof in the logic $\GRPL$ is  a finite sequence of graded formulas
such that each element is either an axiom (A) -- (D) or follows using the rules (E), (F) from
preceding elements  in the sequence. Analogously, a graded proof from a fuzzy theory $\Gamma$
is a finite sequence where each element is an axiom (A) -- (D), or an axiom from the theory 
$\Gamma$ (a graded formula $\gform{q}{\phi}$ where $\Gamma(\phi)=q$),
or follows from its predecessors using the rules (E), (F). 
Clearly, $\GRPL$ is a logic of graded formulas only;
there is no way to derive a non-graded formula.
In case $\gform{r}{\phi}$ is the last element of some proof (possibly from a theory $\Gamma$), 
we say that the formula $\varphi$ has a proof of value $r$ (from $\Gamma$); we write
$\Gamma\vdash \gform{r}{\phi}$ (or just $\vdash\gform{r}{\phi}$ in case $\Gamma$ is empty). 
Finally, $\Gamma\vdash_r \phi$ provided that $r = \sup \{ a \mid 
\Gamma\vdash \gform{a}{\phi} \}$; this is the \emph{provability degree} of $\phi$ in $\Gamma$.

\begin{theorem}{\bf{(Pavelka completeness for $\GRPL$)}}{\rm \cite{Hajek:ArHierarchyI}}
For any fuzzy theory $\Gamma$ and any formula $\phi$,
the provability degree of $\phi$ from $\Gamma$  in $\GRPL$ coincides with its validity degree under $\Gamma$.
\end{theorem}

H\'ajek's paper \cite{Hajek:ArHierarchyI} only presents
the system $\GRPL$ briefly, on its way to an undecidability result (the paper comprises five pages of
two-column print). It is fair to conjecture that if the system were developed in more detail,
it would also have included rules allowing to switch between graded and non-graded formulas. 
Indeed, while Pavelka considers $\alg{L}$-consequence relations, where
$\alg{L}$ stands for a complete lattice (in part I) or a complete $\FLew$-algebra\footnote{Pavelka speaks about complete residuated lattices and this reflects
a mismatch in terminology: ``residuated lattice'' may imply integrality and commutativity (as it did to Ward
and Dilworth, who coined the term in \cite{Ward-Dilworth:ResiduatedLattices}), along with 
boundedness (as it did to Pavelka); or none of these (it is used in this general sense in 
\cite{Galatos-JKO:ResiduatedLattices}).} 
(in part II), and speaks about $\alg{L}$-consequence operators (p.~46) and 
$\alg{L}$-deductive systems, he does also consider usual
formulas in the $\FLew$-langage with constants for elements in some complete $\FLew$-algebra,
and obtains the free algebra from the term algebra, as usual in algebraic logic.

\section{Relationships between the logics $\GRPL$, $\RPL$, and $\Luk$}
\label{section:relationship}

$\RPL$ is an expansion\footnote{A logic $\vdash'$ expands a logic $\vdash$
provided that $\lang{\vdash'}\supseteq\lang{\vdash}$ and for every set of 
formulas $T\cup\{\phi\}$ of $\lang{\vdash}$,
we have $T\vdash\phi$ if and only if $T\vdash'\phi$.} of $\Luk$: 
namely, $\RPL$ is obtained from a presentation of $\Luk$ by adding
the bookkeeping axioms (for all basic connectives used in that presentation). 
Notice that whenever a logic $\logic{L'}$ expands a logic $\logic{L}$, 
this entails conservativity for $\logic{L'}$ over $\logic{L}$.
On the propositional level, the conservativity of $\RPL$ over $\Luk$ 
is an immediate consequence of the finitarity of both logics,
 the local deduction theorem, 
soundness w.r.t.~the standard semantics, 
and standard completeness theorem for $\Luk$; see 
\cite[Prop.~9]{Esteva-Godo-Noguera:RationalWNM} for an analogous consideration in another
logic with constants.
On the other hand, it is nontrivial to prove that first-order Rational Pavelka logic
is conservative over first-order {\L}ukasiewicz logic, where standard completeness is not available;
see \cite[Corollary 2.5]{Hajek-Paris-Shepherdson:Pavelka} for the result.
Thus, $\Luk$ is a fragment of $\RPL$, obtained by considering the language without rational constants
(except the constants $0$ and $1$, either of which are either present in the basic
set of connectives, or term definable from them).

It is clear that $\RPL$ cannot, in turn, be viewed as a fragment of $\L$, 
since $\RPL$ defines functions that $\L$ does not: in particular, if $q$ is a rational number in the interval
$(0,1)$, then there is a term $t$ in the language of $\RPL$ such that the interpretation of $t$ in the 
standard $\RPL$-algebra is the constant function $q$: namely, $t$ is the term $\overline{q}$. 
No such term is available in $\Luk$, due to McNaughton theorem. 

Still, for many purposes,
one can use implicit definability of rationals  in the standard MV-algebra $\standardL$
and in its rational subalgebra $\rationalL$, 
to translate between provable consecutions in $\RPL$ and
provable consecutions in $\L$, and thus effectively ``read'' the former from the latter.
This is based on the account given in \cite{Hajek:1998}, which we proceed to explain.

Consider the algebra $\standardL$.
Let $a\in [0,1]$, let $\phi(x_1,\dots,x_n)$ be a formula in the language $\lang{\Luk}$ 
and consider  an $i$ s.t.~$1\leq i\leq n$. 
The formula $\phi$ \emph{implicitly defines} the element $a$ in variable $x_i$  within $\standardL$  
provided that $\phi$ is satisfiable in $\standardL$ (that is, there is an assignment $v$
such that $v(\phi)=1$ in $\standardL$) and moreover, 
for each assignment $v$ s.t.~$v(\phi)=1$ we have $v(x_i)=a$. 
An element $a$ is (implicitly) definable in $\standardL$ provided that there is a formula that defines it there. 

In other words, an element of the real interval $[0,1]$ is implicitly definable 
provided that it is the only solution, in a given variable, to some finite system of equations in the language
of MV-algebras within $\standardL$. 
Now given a rational $\nicefrac{p}{q}$ and an  $n\in \mathds{N}\setminus\{ 0\}$,
and $i\leq n$, it is easy to define an $n$-variable McNaughton function $f(x_1,\dots,x_n)$ that yields the value $1$ only if $x_i=\nicefrac{p}{q}$: namely
$$f(x_1,\dots,x_n) = 
\begin{cases}
0 \mbox{ if } x_i\leq \nicefrac{(p-1)}{q}\\
qx_i - p + 1 \mbox{ if } x_i \in [\nicefrac{(p-1)}{q}, \nicefrac{p}{q}] \\
-qx_i +p + 1 \mbox{ if } x_i \in [\nicefrac{p}{q}, \nicefrac{(p+1)}{q}]\\
0 \mbox{ if } x_i\geq \nicefrac{(p+1)}{q}.
\end{cases}$$
The function $f$ is term-definable in $\standardL$, so let $\phi_f$ be the formula that defines this function.
Then $\phi_f$ implicitly defines the rational number $\nicefrac{p}{q}$ in variable $x_i$, within
$\standardL$.

Each researcher working in {\L}ukasiewicz logic or MV-algebras may have
their favourite way of defining the rationals in the standard MV-algebra $\standardL$.
Here we rely on a result of Torrens: \cite[Section 2]{Torrens:CyclicElements} implies
that the equation $a^{n-1}=\neg a$ only has one solution in $\standardL$, namely
$\nicefrac{(n-1)}{n}$, and hence the equation $a = (\neg a)^{n-1}$ has the unique solution $\nicefrac{1}{n}$.\footnote{The
latter equation was  also used in \cite[Section 6]{Hanikova-Savicky:SAT}.
With a slight abuse of language, we use \emph{equations} instead of formulas for implicit definitions,
without further mention.}
Under this implicit definition, to define $\nicefrac{m}{n}$ for $m<n$, it is sufficient to consider
the term $m\cdot a$.

We mention in passing that {\L}ukasiewicz logic does not have the Beth property \cite{Hoogland:Thesis}.
Indeed the equivalence $ p\equiv (\neg p)^{n-1}$ implicitly defines $\nicefrac{1}{n}$, while
this constant is not explicitly definable (i.e., term-definable) in $\standardL$ (and consequently, in
$\Luk)$; see also the discussion in \cite[Section 3.3]{Hanikova:ImplicitDefConstLuk}.

On the other hand, if a formula in the language of $\lang{\Luk}$ 
is satisfiable in $\standardL$, then it is satisfied by an assignment in
\emph{rational numbers} (with small denominators) 
\cite{Mundici:Satisfiability,Aguzzoli-Ciabattoni:Finiteness,Aguzzoli:asymptoticBoundLuk,
Hanikova:Handbook}.
It follows that no irrational number is implicitly definable in $\standardL$.\footnote{In infinitary
{\L}ukasiewicz logic, the notion of implicit definability naturally  extends to infinite theories,
whereby each irrational number becomes implicitly definable; see \cite[Section 3.4]{Hanikova:ImplicitDefConstLuk}.}

Thus one can implicitly define precisely each of the rationals in $[0,1]$. 
H\'ajek uses this fact as a technical device to prove finite strong standard completeness of $\RPL$: 
 see \cite[3.3.11 -- 3.3.14]{Hajek:1998}.\footnote{The proof is based on the fact
that, given a  finite set $T\cup\{\phi\}$ of formulas of $\lang{\RPL}$, 
if $ \overline{q_1},\dots,\overline{q_n}$ are all the rational constants in $T\cup\{\phi\}$ and if
$\alpha_1,\dots,\alpha_n$ are formulas that implicitly define $q_1,\dots,q_n$ in $\standardL$
(assuming that the variables occurring in $\alpha_i$ are disjoint from the ordinary propositional
variables and that the variables in $\alpha_i$ and $\alpha_j$ are disjoint whenever $i \not= j$),
then $T\models_{\standardLQ}\phi$ if and only if $T^* \cup\{\alpha_i\}_{i=1}^n \models_{\standardL}\phi^*$,
where $T^*$, $\phi^*$ are obtained from $T$, $\phi$ when each constant $q_i$ is replaced with
the variable in $\alpha_i$ that implicitly defines it---say, a $z_i$.
}
Moreover, as pointed out in \cite{Hanikova:ImplicitDefConstLuk}, 
a straightforward way to obtain an implicit definition of any rational number $\nicefrac{p}{q}$
is via the bookkeeping axioms:
to implicitly define  $\nicefrac{p}{q}$, take a version of the bookkeeping axioms  
and select all such axioms that only  use the constants $\overline{0}, \overline{\nicefrac{1}{q}}, \dots,
\overline{\nicefrac{(q-1)}{q}}, \overline{1}$; call this set $\mathrm{B_q}$.
Notice that this set is finite.
Let $\mathrm{B}_q^z$ be obtained from $\mathrm{B_q}$ by replacing each constant $\overline{\nicefrac{p}{q}}$
with a variable $z_{\nicefrac{p}{q}}$, distinct from ordinary propositional variables, and in such a way that for $p\not=p'$
the variables $z_{\nicefrac{p}{q}}$ and $z_{\nicefrac{p'}{q}}$ are distinct. 
Notice that $\mathrm{B}^z$ features no propositional constants
except possibly $0$ and $1$.
Then  
the theory $\mathrm{B}_q^z$ (is consistent and) implicitly defines each of the rationals 
$0, \nicefrac{1}{q},\dots,\nicefrac{(q-1)}{q},1$.
In particular, $\mathrm{B}_q^z$ proves the Torrens formula 
$z_{\nicefrac{1}{q}} \equiv (\neg z_{\nicefrac{1}{q}})^{q-1}$ in $\Luk$. 
If $v$ is a model of $\mathrm{B}_q^z$, we have
$v(z_{\nicefrac{p}{q}}) = \overline{\nicefrac{p}{q}}$ for each  $q\in\mathds{Q}$. 
Finally, this construction is independent of the choice of the set of bookeeping axioms
(which, in turn, depends on the set of basic connectives of $\Luk$).

It follows that the only expansion of the algebra $\standardL$ with rational constants 
is the \emph{canonical} one, i.e., the one where the algebraic constant 
$\overline{q}$ is interpreted as $q\in[0,1]$. 
This was  proved in \cite{Esteva-GGN:AddingTruthConstants} (see 
the remark under Theorem 26 of that paper). 
This is in tight relation to the fact that the standard MV-algebra has no nontrivial 
automorphisms \cite[Corollary 7.2.6]{Cignoli-Ottaviano-Mundici:AlgebraicFoundations}.
This fact may render the notion ``canonical'' rather 
superfluous in $\RPL$; but the notion becomes important in, e.g., 
the expansion of product logic with constants, where not all interpretations
of constants in the standard product algebra are canonical; see \cite{Savicky-CEGN:ProductTruthConstants}.

We now proceed to examine the relationship between the logics $\RPL$ and $\GRPL$.
Given how H\'ajek's works around these topics (namely, 
the papers \cite{Hajek:ArHierarchyI,Hajek:FuzzyLogicPointOfView} and 
 finally, section 3.3 in \cite{Hajek:1998}) are presented,  
it can be assumed that the series of lemmas here were known to him. 
Technically the statements are very easy to derive 
provided only one is familiar with both the systems $\RPL$ and $\GRPL$.
We present them here so that the reader may appreciate some of their implications,
which bear on the discussion of the respective merits of fuzzy logics with graded syntax
vs.~those with classical syntax. 


\begin{lemma}
\label{push_pull}
The following are derivable rules in $\GRPL$:
$\gform{q}{\phi} \triangleleft\,\triangleright\, \gform{1}{\overline{q}\to\phi}$. 
\end{lemma}
\begin{proof}Use graded modus ponens for right-to-left direction; lifting for the converse one.
\end{proof}

\begin{lemma} 
\label{GRPL_in_RPL}
{\rm ($\GRPL$ embeds in $\RPL$)}
Let $T$ be a set of graded formulas and $\gform{q}{\phi}$ a graded formula.
Then $T\vdash_{\GRPL}\gform{q}{\phi}$ if and only if
 $\{\overline{r}\to\psi \mid \gform{r}{\psi}\in T\} \vdash_{\RPL} \overline{q}\to\phi$.
\end{lemma}

\begin{proof}
First assume $T\vdash_{\GRPL}\gform{q}{\phi}$:
 there is a sequence of graded formulas such that each element is an instance of an axiom of $\GRPL$, 
a $T$-axiom, or obtained from earlier elements using graded modus ponens or lifting,
and the last element is $\gform{q}{\phi}$.
From this sequence, the desired proof in $\RPL$ can easily be obtained with the help of 
the following observations:\\
-- if $\gform{q}{\phi}$ is an axiom instance of $\GRPL$, then $\overline{q}\to\phi$ is a theorem of $\RPL$;\\
-- the rules \ 
\AxiomC{$\overline{r}\to {\varphi}$}
\AxiomC{$\overline{s}\to{\phi\to\psi} $}
\BinaryInfC{$\overline{r\odot^\mathrmL s}\to{\psi}$}
\DisplayProof \ 
and \ 
\AxiomC{$\overline{q}\to\phi$}
\UnaryInfC{$(\overline{r \to^\mathrmL q}) \to (\overline{r}\to\phi)$}
\DisplayProof \  are derivable in $\GRPL$.
The rule on the left is Lemma 3.3.2 in \cite{Hajek:1998} (still named ``graded modus ponens'' in that text).
The one on the right is provable thanks to the provability of the implication 
$(\overline{q}\to\phi) \to ((\overline{r}\to\overline{q}) \to (\overline{r}\to\phi))$
(an instance of the  axiom (A1) of H\'ajek's Basic Logic BL, cf.~\cite{Hajek:1998}) 
in the language of $\RPL$; 
then swap $\overline{r}\to\overline{s}$ for $\overline{r\to^\mathrmL s}$  using bookkeeping.

Assume now that the sequence $A = \alpha_1,\dots,\alpha_n$ is an $\RPL$-proof of 
$\overline{q}\to\phi$ from $\{\overline{r}\to\psi \mid \gform{r}{\psi}\in T\}$:
 each $\alpha_i$ is an axiom instance of $\RPL$, a formula $\overline{r}\to\psi$ for $\gform{r}{\psi}$ 
an element of $T$, or derived from earlier elements of $A$ using modus ponens,
while $\alpha_n$ is $\overline{q}\to\phi$.
Take the sequence of graded formulas $A' = {\alpha'}_1,\dots,{\alpha'}_n$ with ${\alpha'}_i = \gform{1}{\alpha_i}$.
Notice that if $\alpha_i$ is provable from some $\alpha_j$ and $\alpha_k$ (with $ j,k <i$) in $\RPL$,
then ${\alpha'}_i$ is provable from ${\alpha'}_j$ and ${\alpha'}_k$ using graded modus ponens in $\GRPL$.
If $\alpha_i$ is an axiom instance of $\RPL$, then $\gform{1}{\alpha_i}$ is an axiom instance of $\GRPL$,
and if $\alpha_i$ is $\overline{r}\to\psi$ for $\gform{r}{\psi}\in T$, then $\gform{1}{\alpha_i}$ and
$\gform{r}{\psi}$ form an invertible rule in $\GRPL$, by Lemma \ref{push_pull}. 
Hence we have a proof of ${\alpha'}_n$ in $\GRPL$ from $T$; using again 
Lemma \ref{push_pull} to get $\gform{q}{\phi}$ from ${\alpha'}_n$ in $\GRPL$, 
one obtains the desired $\GRPL$-proof.
\end{proof}

The following lemma is immediate:

\begin{lemma}
\label{lm_RPL_proofs_grade_1}
Let $T$ be a set, and $\phi_1, \phi_2, \dots, \phi_n$ a sequence,
  of formulas in  $\lang{\RPL}$.
Then $\phi_1, \phi_2, \dots, \phi_n$ is a proof in $\RPL$ from $T$ if and only if
the sequence $\gform{1}{\phi_1}$, $\gform{1}{\phi_2}$, \dots, $\gform{1}{\phi_n}$ (all grades $1$)
is a proof in $\GRPL$ from $\{\gform{1}{\psi}\mid \psi\in T\}$. 
\end{lemma}

\begin{lemma}
\label{proof_all_1}
Let $T_0$ be a set of formulas in $\lang{\RPL}$, $T$ the set of graded formulas
$\{\gform{1}{\phi}\mid \phi\in T_0\}$, and  $\gform{1}{\phi}$ a graded formula.
Assume  $T\vdash_{\GRPL} \gform{1}{\phi}$.
Then there is a $\GRPL$-proof $P$ of $\gform{1}{\phi}$ from  $T$
such that all formulas in $P$ have the grade $1$.
\end{lemma}

\begin{proof}
Let $P_0$ be any $\GRPL$-proof of $\gform{1}{\phi}$  from $T$: a sequence
 $\gform{q_1}{\phi_1},\dots,\gform{q_n}{\phi_n}$,  such that $q_n$ is $1$ and $\phi_n$ is $\phi$.

Define a sequence $P_1$ of $n$ graded formulas: 
if $\gform{q_i}{\phi_i}$ is the $i$-th element in $P_0$, 
let $\gform{1}{\overline{q_i}\to\phi_i}$ be the $i$-th element in $P_1$. 

Now recall that if $\gform{r}{\varphi}$ is an axiom of $\GRPL$,
then $\gform{1}{\overline{r}\to\varphi}$ is provable in $\GRPL$;
and moreover, analogously to the observations in proof of Lemma \ref{GRPL_in_RPL}, 
 the rules \\ 
\AxiomC{$ \gform{1}{\overline{r} \to \phi}$}
\AxiomC{$\gform{1}{\overline{s}\to (\varphi\to\psi)} $}
\BinaryInfC{$\gform{1}{\overline{r\odot^\mathrmL s}\to \psi}$}
\DisplayProof \ 
and \ 
\AxiomC{$ \gform{1}{\overline{r}\to\phi}$}
\UnaryInfC{$\gform{1}{(\overline{s}\to \overline{r}) \to (\overline{s}\to \phi)}$}
\DisplayProof \ 
are derivable in $\GRPL$.

But $P_1$ is not a proof from $T$: if $\gform{1}{\psi}$ is an element of the sequence $P_0$
for some formula $\psi\in T_0$, this has been replaced in $P_1$ by the formula $\gform{1}{1\to\psi}$.
But these two graded formulas are interderivable by Lemma \ref{push_pull},
so one can formally prefix the sequence $P_1$ with the finitely many formulas
 $\gform{1}{\psi}$ from $T$ that are used in the proof.
This (possibly longer) sequence $P_2$ is indeed a $\GRPL$-proof from $T$.  
Finally, one derives $\gform{1}{\phi}$ from the last line $\gform{1}{\overline{1}\to\phi}$ of $P_2$,
using Lemma \ref{push_pull}. This yields the sequence $P$.
\end{proof}

\begin{corollary}
\label{RPL_in_GRPL}
{\rm ($\RPL$ embeds in $\GRPL$.)}
Let $T\cup\{\phi\}$ be a set of formulas of the language of $\RPL$.
Then $T\vdash_{\RPL}\phi$ if and only if
$\{ \gform{1}{\psi}\mid \psi\in T\} \vdash_{\GRPL}\gform{1}{\phi}$.
 \end{corollary}

\begin{proof}
Left to right: use Lemma \ref{lm_RPL_proofs_grade_1}. Right to left: use Lemma \ref{proof_all_1},
then again apply Lemma \ref{lm_RPL_proofs_grade_1}.
\end{proof}

\begin{corollary}
\label{GRPL_in_GRPL}
{\rm ($\GRPL$ embeds in $\GRPL$.)}
Let $T$ be a set of graded formulas and $\gform{q}{\phi}$ a graded formula.
Then $T\vdash_{\GRPL} \gform{q}{\phi}$  if and only if $\{ \gform{1}{r\to\psi}\mid\gform{r}{\psi}\in T\}
\vdash_{\GRPL}\gform{1}{q\to\phi}$. Moreover, the proof whose existence is asserted on the right
can be chosen in such a way that all graded formulas in the proof have the grade $1$.
\end{corollary}

In plain words, the provability relation in $\GRPL$ is fully captured by those formulas and proofs
in the logic $\GRPL$ whose degree is $1$. 
This is witnessed by an embedding that can be carried out with minimal overhead.

\section{Discussion}
\label{section:discussion}

Several authors, on various occasions, voiced their views on the respective similarities 
and differences between the three types of logics of our interest. 
An early example among such remarks is the already quoted one (see Sections \ref{section:intro},\ref{section:evo}) of 
Goguen \cite[p.~365]{Goguen:LogicInexactConcepts} on semantics making its way into the syntax;
on this issue, recall our discussion of $\RPL$-algebras as expansions of MV-algebras in Section \ref{section:formally},
pointing out that the introduction of constants and the bookkeeping axioms 
significantly restrict the range of algebras that form the equivalent algebraic semantics.
This is peculiar to MV-algebras, for some other classes (such as product algebras) the situation may be different.

Interestingly, H\'ajek offered subsequently two comments that may indicate some evolution of his view.
The earlier one can be found in  \cite{Hajek:FuzzyLogicPointOfView} and pertains to $\RPL$\footnote{The numbering 
of the references in this and subsequent quotations is replaced with the numbering of the same works
in our references.}: 
``Logics of partial truth were studied, in a very general manner, as early as in the seventies by the Czech mathematician Jan Pavelka
\cite{Pavelka:Fuzzy} and since then have been substantially simplified; I refer to \cite{Hajek:ArHierarchyI} but here we describe
a still simpler version. It is very different from the original Pavelka's version and looks as an `innocent' extension of {\L}ukasiewicz's $\Luk$;
but the main completeness result of Pavelka still holds.''
The paper  \cite{Hajek:FuzzyLogicPointOfView} is a survey of various systems of fuzzy logic, presented at the SOFSEM conference
to an audience of varied background; it does not give any proofs. Nevertheless, 
it is within this paper that H\'ajek first introduces $\RPL$ in the form in which it will have, some years later, been presented in \cite{Hajek:1998} in full.
Compared to the presentation of \cite{Hajek:ArHierarchyI}, it omits graded formulas altogether, 
while both papers employ constant only for the rational elements, and both simplify the axioms.
We stress that the comparison in  \cite{Hajek:FuzzyLogicPointOfView} is made between $\RPL$ on the one hand, 
and (not $\GRPL$ but) Pavelka's logic from \cite{Pavelka:Fuzzy} on the other; thus indeed there are many differences.
But our view is that the ``innocence'' mentioned in the quotation refers predominantly to the absence of graded formulas in $\RPL$, rather than the other (notable) differences such as the countability of language of $\RPL$ and simpler axiomatization for the $\Luk$-fragment.
By the time his monograph on fuzzy logic was finished, though, H\'ajek had settled into the following view  \cite[Section 3.3]{Hajek:1998}:
``A \emph{graded formula} is a pair $(\phi,r)$ where $\phi$ is a formula and $r$ a rational element of $[0,1]$; it is just another notation for
the formula $(\overline{r}\to\phi)$.'' In other words, H\'ajek takes the embedding we spelled out in Lemma \ref{GRPL_in_RPL} for granted,
and considers the difference between graded formulas and implications to be \emph{merely notational}. 

Some authors  do not really distinguish between
the logics that employ graded formulas explicitly (here represented by the logic $\GRPL$) 
those that merely employ constants (such as $\RPL$):
in \cite[p.~628]{Esteva-Godo-Marchioni:FuzzyLogicsEnriched} Esteva, Godo, and Marchioni write: 
``On the other hand, in some situations one might be also interested in explicitly
representing and reasoning with intermediate degrees of truth. A way to do so, while
keeping the truth preserving framework, is to introduce truth-constants into the language. This 
approach actually goes back to Pavelka [75], who built a propositional many-valued
logical system which turned out to be equivalent to the expansion of {\L}ukasiewicz logic
obtained by adding to the language a truth-constant $r$ for each real $r \in [0, 1]$, together
with some additional axioms.'' Here, of course, the two equivalent systems
both feature uncountable set of constants, but the difference between them is 
of the same nature as with $\GRPL$ and $\RPL$. The equivalence in this quotation
is rendered in Section \ref{section:relationship} in Lemmas \ref{GRPL_in_RPL} and \ref{RPL_in_GRPL}.

Finally, let us recall the opinion H\'ajek, Paris, and Shepherdson \cite{Hajek-Paris-Shepherdson:Pavelka}, 
the paper which provided a cue for the title of our survey.\footnote{The numbers of theorems in the quotation 
refer to the numbering in \cite{Hajek-Paris-Shepherdson:Pavelka}, and the text in square brackets is a part
of the original quotation, rather than an insertion of the current author.} ``From our conservation results it could be argued that Rational Pavelka logic is the preferred system to use in that it has the best of both worlds.
It does not extend {\L}ukasiewicz logic (Theorem 2), even for statements involving partial truth (Theorem 2.6 and Section 3), yet adding truth constants
for all real truth values in $[0,1]$ (to give the original Pavelka logic) only produces a conservative extension (Theorem 2.7). [On the other hand one might
also argue that since the two logics are equivalent, {\L}ukasiewicz logic is to be preferred because of its greater syntactical simplicity.]''
Here, the two worlds that the quotation refers to appear to be {\L}ukasiewicz logic on the one hand and Pavelka logic on the other;
thus $\RPL$ is neither, and is viewed as a suitable compromise between the two. 
The equivalence stipulated within the brackets is accounted for by Theorem 2.6 of the paper.

These views of H\'ajek and other authors certainly have met with considerable dissent.
It turns out that it has been the use of graded formulas, theories, and proofs, 
rather than a mere presence of  rational (or real) constants, 
 what many researchers have taken to be the genuine insignia of fuzzy logic.
And moreover, since the graded logic of Pavelka, developed in the late 1970's, preceded the tumultuous development of mathematical fuzzy logic at the turn of the century, it has been the involvement of H\'ajek
that has been perceived as a dissent.

One such stance was expressed in the work of B\v{e}lohl\'avek, while  analyzing the wake of Pavelka's work,
in his survey paper \cite{Belohlavek:PavelkaRetrospectProspect} ``This $\RPL$ [$\GRPL$ in our sense] is still a kind of abstract fuzzy logic with the notion of degree of provability
defined as by Pavelka.//
He [H\'ajek] later presented $\RPL$ in his \cite{Hajek:1998}. This $\RPL$, 
however, is conceptually different from the $\RPL$ of \cite{Hajek:ArHierarchyI} [$\GRPL$ in our sense]. 
Namely, it is not an abstract fuzzy logic but rather an expansion of the ordinary {\L}ukasiewicz 
infinitely-valued logic by truth constants for rationals with extra axioms 
regarding truth degrees (essentially same as those described above). In
this logic, the genuine notions of abstract fuzzy logic, 
such as that of degree of provability, are `simulated' by the
ordinary notions.'' B\v elohl\'avek makes a direct comparison of the logics $\RPL$ and $\GRPL$, 
pinpointing the perceived difference (which rests neither in the cardinality of the set of constants, nor in the
axiomatization of the $\Luk$-fragment, but in the presence of graded formulas and graded proofs and the 
 natural way in which they yield the notion of provability degree.

Still more dismissive views on of H\'ajek's simplifications of Pavelka logic, which eventually yielded $\RPL$, 
have been put forward by Nov\'ak \cite[p.~1100]{Novak:Handbook}:
``Then the \emph{provability degree} of a formula $A$ is defined [\dots] Unfortunately, H\'ajek did not provide a sound justification
of this notion and introduced it only as an additional and not really organic concept. $\RPL$ thus became a special extension of {\L}ukasiewicz logic,
for which it is unclear why it should be studied.''

In this survey, we have taken some trouble to spell out simple technical lemmas that 
other authors may gloss over in pursuit of more interesting mathematics.
The technicalities hopefully help to make the following simple point. 
The transition from $\GRPL$ to $\RPL$ begins by admitting the validity of our Corollary \ref{GRPL_in_GRPL}: 
 the logic $\GRPL$ embeds into itself in such a way that
all formulas in the range of the embedding have degree $1$. Moreover, each $\GRPL$-provable consecution on a set of premises with all grades $1$ and a conclusion with grade $1$
has a $\GRPL$-proof in which each formula also has grade $1$.
The embedding is obtained invoking the lifting rule, present both in $\GRPL$
and in Pavelka logic.

Even if the logic $\RPL$ had never been conceived and developed, as a next step in the
investigation of fuzzy logics with constants,
it would still be clear, from the metamathematical considerations on $\GRPL$ alone, that
 extant grades for all formulas that the formalism employs are an explicit formulation of 
what is implicit already in the fragment of $\GRPL$
that only uses the grade $1$: this includes provability degrees and (a version of) Pavelka completeness.
Put yet differently, $\GRPL$ \emph{simulates itself} (in the same sense in which B\v{e}lohl\'avek claims $\RPL$ simplates $\GRPL$) with graded formulas where all grades are $1$. 

Given the above, our inquiry is not inconclusive: 
we argue the view that graded syntax systems, such as $\GRPL$, merged into the traditional ones, such as $\RPL$,
due to an application of Occam's razor, 
which yielded an ostensibly simpler, algebraizable logic, with fewer axioms, 
with the usual syntax and a few truth constants thrown in.
The fact that the provability relation in $\GRPL$ can be fully encoded in formulas and proofs with all grades 
$1$ makes the need for graded formulas questionable. In particular,
 researchers championing graded syntax admit that
it hardly makes sense to explicitly employ grades if they are all $1$: ``It should be noted that traditional syntax can also be taken
as evaluated [graded in our sense], but the only evaluations of formulas are $1$ or $0$. This is an extreme situation and of course,
it can hardly have any sense to write explicitly the evaluation $1$ with each formula.'' (see \cite[p.~1063]{Novak:Handbook})
The introduction of the logic $\RPL$ thus did precisely this: dropped the degree $1$ and 
 made more explicit the above mentioned embedding. 

We pointed out that $\RPL$ restricts the class of algebras that forms the equivalent algebraic semantics
in comparison to $\Luk$: only the trivial MV-algebra or an  MV-algebra that contains an isomorphic copy of the MV-algebra on the rationals in $[0,1]$ with the usual order as a subalgebra can be expanded to an $\RPL$-algebra.
Despite this fact, are there still reasons to prefer $\RPL$ over $\Luk$?
It is useful to realize here that already the MV-algebra on the rationals in $[0,1]$ with the usual order has
the \emph{desiderata} for graded reasoning in a finite setting (which covers most real-life situations,
except those where infinite models of appear, for some reason, preferable): bounded, dense, totally  ordered.
In fact for such reasoning in a finite setting, a finite MV-chain is sufficient, and
the MV-chain on the rationals is the smallest MV-chain among those MV-algebras into which all finite MV-chains embed.
To conclude, we borrow an elegant quotation from \cite[p.~677]{Hajek-Paris-Shepherdson:Pavelka}:
``\dots even for
partial truth, Rational Pavelka logic deals with exactly the same logic as {\L}ukasiewicz
logic---but in a very much more convenient way.''
The elegance is a cognitive one---we remark in passing, for example, that it is computationally hard
to decide which rational value (if any) is implicitly defined by a particular formula of {\L}ukasiewicz logic
\cite[Lemma 4.6]{Hanikova:ImplicitDefConstLuk}.

\section{Concluding remarks}
\label{section:conclusion}

The material presented in this survey can be read as a contribution,
within the family of {\L}ukasiewicz logic, to the study of fragments of logics,
embeddings between logics, and the merit of studying these phenomena.
It is clear  that either of the embeddings between $\RPL$ and $\GRPL$ 
is computationally feasible, in fact almost trivial.
The same is true---but less obviously---about translating the finite provability relation of $\RPL$ into 
the finite provability relation in $\Luk$; this was shown in \cite{Hajek:1998,
Hajek-Paris-Shepherdson:Pavelka}, moreover  by \cite[Theorem 4.3]{Hanikova:ImplicitDefConstLuk}
it can be done via a function operating in polynomial time in the size of the input.

One direction for a future research in the area of expanded languages 
is indicated by the close connection between MV-algebras and lattice-ordered abelian groups:
since the variety is generated by the multiplicative group on the positive rationals,
it is natural to wonder about the relationship between the theory of this group and 
the theory of its its expansion with constants for rationals.

\bigskip
{\bf Acknowledgements.}
This work was supported partly by the grant GA18-00113S of the Czech Science Foundation
 and partly by the long-term strategic development financing 
of the Institute of Computer Science (RVO:67985807).
The author would like to thank Carles Noguera for reading and commenting upon a draft of the manuscript.

\bibliographystyle{plain}

\end{document}